\documentstyle[12pt,aasms4]{article}
\slugcomment{}
\begin{document}
\title{Peaks in the Cosmic Microwave Background: flat versus open models} 
\author{R. B. Barreiro\altaffilmark{1}, J. L. Sanz, E. Mart\'\i nez-Gonz\'alez}
\affil{Instituto de F\' \i sica de Cantabria, CSIC-Universidad de Cantabria,\\
Facultad de Ciencias, Avda. Los Castros s/n, 39005 Santander, Spain}           
\and 
\author{L. Cay\'on, Joseph Silk}  
\affil{Astronomy Department and Center for Particle Astrophysics, \\
University of California, Berkeley, CA 94720}
\altaffiltext{1}{Dpto. de F\' \i sica Moderna, Universidad de Cantabria,
Facultad de Ciencias, Avda. Los Castros s/n, 39005 Santander, 
Spain}  
\begin{abstract}

We present properties of the peaks (maxima) of the microwave background
anisotropies expected in flat and open cold dark matter models. 
We obtain analytical expressions of several topological descriptors: mean
number of maxima and the probabilty distribution of the gaussian curvature and
the eccentricity of the peaks. These quantities are calculated as functions of
the radiation power spectrum, assuming a gaussian distribution of temperature
anisotropies.
We present results for angular resolutions ranging from $5'$ to 
$20'$ (antenna FWHM),
scales that are relevant for the MAP and COBRAS/SAMBA space missions and the 
ground-based interferometer experiments. 
Our  analysis also includes 
the effects of noise. We find that the number of peaks can discriminate 
between standard
CDM models, and that the gaussian curvature distribution provides 
a useful test for these 
various models, whereas the eccentricity distribution can not distinguish 
between them.

\end{abstract}


\section{Introduction}\label{intr}

The standard way to study the microwave background anisotropies (CMB) is based 
on the computation of the radiation power spectrum, i.e., the $C_\ell's$. 
The texture 
of the CMB offers a useful alternative to this approach and can be used to 
test 
models of galaxy formation. Let us consider the excursions of a gaussian
random field above a certain threshold $\nu = (\Delta T)/(\Delta T)_{rms}$. It
is of interest to remark that once $\nu $ is fixed, all the topological 
quantities we will calculate are normalization--independent.
Earlier work on the properties of peaks in one-dimensional scans and 2D maps of
the CMB are due to 
Zabotin \& Nasel'skii (1985) and Sazhin (1985), respectively. A key paper on 
two-dimensional
fields and its implications for the CMB is that of  Bond \& Efstathiou (1987).
This technique was applied to  calculate the number of spots on small angular 
scales in different 
models (Vittorio \& Juszkiewicz 1987; Mart\'\i nez-Gonz\'alez and Sanz
1989) and to study the Tenerife experiment (Guti\'errez et al. 1994). 
A similar analysis applied to non-gaussian random fields have been 
performed by Coles \& Barrow (1987). 

   In this Paper, we consider the statistical properties of the CMB,
assuming that the temperature fluctuations can be represented by 
a two-dimensional gaussian random field. The local description of maxima is
presented in \S2. We
will restrict our analysis to the peaks of the field above a certain
threshold. In particular we are interested in the following quantities: mean
number of peaks over the whole celestial sphere $N(>\nu )$ (\S3), 
gaussian curvature 
probability density function (p.d.f.) $p(\kappa , >\nu )$ (\S4) 
and eccentricity p.d.f.
$p(\epsilon , >\nu )$ (\S5). All of them can be calculated analytically 
in terms of  parameters that are related to the CMB radiation power spectrum.
This power spectrum is characterized by the Doppler peaks and a cut-off at high
$\ell$, which depend on the cosmological parameters (for a recent review see
Hu 1996). An accurate determination
of the radiation power spectrum requires detailed numerical calculations in
perturbation theory (Sugiyama 1996).
Flat and open cold dark matter models
with a Harrison-Zel'dovich initial power spectrum will be considered. 
Our analysis includes an angular resolution 
ranging from $5'$ to $20'$, of interest for future experiments, and also
the effects of noise.
Conclusions are presented in \S6.
                                                              
\section{Local description of maxima}\label{local}

The local description of maxima involves the second derivatives of the field
along the two principal directions. As usual, the curvature radii are defined 
by:
$R_1 = [-\Delta ^{\prime \prime }_1(max)/2]^{-1/2}$ and 
$R_2 = [-\Delta ^{\prime \prime }_2(max)/2]^{-1/2}$,
where $\Delta $ is the temperature field normalized to the rms-fluctuations.
Then  with any maximum, we can associate  two invariant quantities: the 
gaussian curvature $\kappa $ and the eccentricity $\epsilon $ given by

\begin{equation}
\kappa = \frac {1}{R_1R_2},\ \ \ \epsilon = \left[1 -
\left(\frac{R_2}{R_1}\right)^2\right]^{1/2}.
\end{equation}

The number density of peaks of a two-dimensional homogeneous and isotropic
gaussian random field has been studied by Longuet-Higgins (1957) and Bond 
and Efstathiou (1987). After a
straightforward calculation, one can obtain the mean number of maxima (over the
celestial sphere) $N(\kappa ,\epsilon ,\nu )d\kappa \,d\epsilon \,d\nu $ with 
gaussian curvature, eccentricity and threshold between ($\kappa ,
\kappa + d\kappa $), ($\epsilon, \epsilon + d\epsilon $) and ($\nu, \nu + 
d\nu$), 
respectively. $N$ is given in terms of two spectral parameters
$\gamma$ and $\theta_*$ that charaterize the intrinsic cosmological model plus
the noise

\begin{equation}
\theta_c = 2^{1/2}\frac{\sigma _0}{\sigma _1},\ \ \ 
\gamma = \frac{\sigma_1^2}{\sigma_0 \sigma_2},\ \ \ 
\theta_* = 2^{1/2}\frac{\sigma _1}{\sigma _2},\ \ \
\theta_* = \gamma \theta _c ,
\end{equation}
\begin{equation} 
\sigma_0^2 = C(0,\sigma ),\ \ \ 
\sigma_1^2 = -2\,C^{\prime \prime }(0,\sigma ),\ \ \   
\sigma_2^2 = \frac{8}{3}\,C^{(iv)}(0,\sigma ),
\end{equation}

\noindent where $\sigma $ is the gaussian dispersion ($\sigma=0.425\times 
FWHM$). 
The $C_\ell$'s have two different contributions: the intrinsic cosmological 
signal and the noise. We will consider flat and open CDM models
($\Omega = 1, 0.3, 0.1$, baryon content $\Omega _b = 0.05$, Hubble constant $h
= 0.5$) with adiabatic fluctuations and a Harrison-Zel'dovich primordial 
spectrum, kindly provided by N. Sugiyama. The $C_\ell$'s
have been normalized to the COBE 2-year maps (Cay\'on et al. 1996. This
normalization does not appreciably change with the 4-year data). 
We consider several noise amplitudes assuming that all scales contribute at the
same level (white noise). The multipole coefficients of the noise $C_{(N)}$ 
are therefore given by:

\begin{equation} 
C_{(N)} = A_{(N)}^2(10')\frac{4\pi}{\sum _\ell (2\ell + 1)
e^{-\ell(\ell + 1)\sigma ^2}}\ \ \ . 
\end{equation}

\noindent
The noise amplitude $A_{(N)}(10')$, i.e. the noise after smoothing with 
a $10'$ FWHM gaussian window, is fixed at $A_{(N)}(10')=(0,1,3)\times 10^{-5}$
with $\sigma=0.425\times 10'$, giving $C_{(N)}=(0,1.9,17)\times 10^{-15}$ for
the three noise levels used in our examples
(a justification will be given below). Then, $A_{(N)}$ for other 
angular resolutions (5$'$,20$'$) can be
obtained using the same $C_{(N)}$ through the previous formula.

Following standard observational procedures, we will filter signal plus noise 
with a gaussian with approximately the same width than the antenna FWHM.
The angular correlation function $C(\alpha ,\sigma )$ is therefore given by

\begin{equation}   
C(\alpha , \sigma ) = \frac{1}{4\pi }\sum _\ell (2\ell + 1)
\left(C_\ell e^{-\ell(\ell + 1)\sigma ^2}+C_{(N)}\right) 
P_\ell (\cos \alpha )e^{-\ell(\ell + 1)\sigma ^2},\ \ \ 
\end{equation}

In Table 1 we give the coherence length and the parameters $\gamma $ and
${\theta }_*$ for $\Omega =0.1, 0.3, 1$, different angular resolutions and
$A_{(N)}(10')$'s.
$\theta_*$ increases with beam size and decreases with $A_{(N)}(10')$. 
When no noise is
present the coherence angle has a range between $8'.6$ and $35'.9$ for the   
values of
the parameters considered. That range decreases as the noise level increases.

We will analyse 2D temperature fluctuations (signal plus noise) with 
angular resolution $FWHM(') = 5, 10$ and $20$, 
which are of interest for the most sensitive bolometers and radiometers of 
future space experiments (COBRAS/SAMBA and MAP) and also for the VSA 
experiment as well as other interferometric experiments. The values of
$A_{(N)}(10')$ considered in this paper cover the range of sensitivities
expected for the future experiments.
In particular, the best expected sensitivity of COBRAS/SAMBA corresponds, in
practice, to the case $A_{(N)}(10')=0$.

\section {Number of peaks}

The number of peaks above the threshold $\nu $, $N(>\nu )$, can be calculated
from the differential number $N(\nu )d\nu $

\begin{eqnarray}
N(\nu ) & = & N_T \left(\frac{6}{\pi }\right)^{1/2}e^{-{\nu}^2/2}\left[
 {\gamma }^2 
\left({\nu}^2-1\right)\left(1 - \frac{1}{2} {\rm erfc }(\gamma \nu s)\right) + 
\right.\nonumber\\
 & &\left. \nu \gamma 
\left(1 - {\gamma }^2\right)\frac{s}{{\pi }^{1/2}}e^{-{\gamma }^2{\nu }^2s^2} +
t\left(1 - \frac{1}{2}{\rm erfc }(\gamma \nu st)\right)e^{-{\gamma }^2{\nu }^2t^2}
\right],
\end{eqnarray}

\begin{equation}
s = \left[2\left(1 - {\gamma }^2\right)\right]^{-1/2},\ \ \  
t = \left(3 - 2{\gamma }^2\right)^{-1/2} 
\end{equation}

\noindent and $N_T = \left(3^{1/2}{\theta _*}^2\right)^{-1}$ is the total number 
of peaks 
over the whole celestial sphere.

In Figure 1, we show the cumulative number of peaks for different 
values of
$\Omega $ and $A_{(N)}(10')$. Generically, the number of peaks increases 
if we
decrease either the beam size or the $\Omega $ parameter, except for $FWHM=5'$
with $A_{(N)}(10')=3\times 10^{-5}$ and
$FWHM = 20'$ where the number is greater for $\Omega = 0.3$. 
For a noiseless map (i.e. $A_{(N)}(10')=0$) and angular resolution of 
$5'$, the number of peaks above the threshold
$\nu=3$ for $\Omega = 0.1$ is approximately 3 times the value
 for $\Omega = 1$ 
(i.e.
4541 as compared to 1657 peaks for the open and flat cases, respectively). 
However, when noise is present, the most favourable case is an angular
resolution of 10$'$, at which the noise decreases considerably while the signal
is slightly affected. In fact,
using a $\chi^{2}$ test and assuming Poissonian errors, the hypothesis 
that the flat and open models are derived from the same
population is rejected at a confidence level $\gtrsim 99\%$
except for an angular 
resolution of $5'$,
$A_{(N)}(10')=3 \times 10^{-5}$ and $\Omega=0.3$. Moreover, as previously
indicated, the best confidence level is attained at the $10'$ angular
resolution.
Since we are considering very small angular scales, the cosmic 
variance will not affect our results from the practical point of view.

On the other hand, we may ask whether gravitational lensing
can change these results. An estimate of the coherence length
including, or not including, lensing  leads 
to $(\theta^{gl}_c/\theta_c)^2\simeq 
1-a^2$, with $a\equiv (\sigma(\beta)/\beta)_{\beta=0}$ being the relative
bending dispersion at zero lag. For standard CDM and low-$\Omega$ CDM models:
$a\lesssim 0.18$, so the number of maxima (which at high thresholds is
 approximately
proportional to the coherence length) is only slightly
modified, i.e. $\lesssim 3\%$ (Mart\'\i nez-Gonz\'alez, Sanz and Cay\'on
1996).

In Table 2, we give the number of peaks above the thresholds $\nu=3,3.5,4$
for different values of $\Omega $, angular resolutions and 
levels of $A_{(N)}(10')$.

	An equivalent quantity that can be used is the mean area of the peaks
above a certain threshold (defined as the total area above that threshold divided by
the corresponding number of peaks). The behaviour of this quantity can be 
easily obtained from the number of peaks and so it does not incorporate any
new information that discriminates between the different models.
As an example, for the case of FWHM=$10'$, $\nu=3$ and
 $A_{(N)}(10')=3\times 10^{-5}$, we find a mean 
area $(arcmin^2)$ of 42.4 for $\Omega=0.1$ and 46.7
for $\Omega$=1, whereas for the $A_{(N)}(10')=0$ these values 
increase to 132.0 and 266.1, respectively.

\section {The distribution of gaussian curvature}

The distribution of peaks above the threshold $\nu$ with inverse of the 
gaussian curvature $L \equiv {\kappa }^{-1}$
between $(L,L+dL )$, $p(L, >\nu)$, can be obtained 
from the following p.d.f. 
                                                                           
\begin{equation}
p(L, \nu ) = \left(\frac 6 \pi \right)^{1/2}a^4tL^{-5}e^{a^2L^{-2}} 
e^{-\frac{3}{2}t^2{\nu }^2}
{\rm erfc}\left[ \frac{s}{t} \left( aL^{-1} - \gamma \nu t^2\right) \right],
\end{equation}
\begin{equation}
a = 2 \gamma {\theta _c}^2
\end{equation}

In Figure 2, we represent the p.d.f $p(L, >\nu)$ for different 
angular
resolutions (FWHM$(') = 5, 10, 20$) and threshold $\nu=3$ . In all the cases, 
except
when the beam size is very small ($5'$) and the noise amplitude $A_{(N)}(10')$ 
high, the curves associated with flat
and open models clearly differ. In those cases, using a KS-test, the null
hypothesis that the flat and open models are derived from the same population
is rejected at a confidence level $\gtrsim 99\%$. The best case is
obtained for an angular resolution of 10$'$ when noise is present. 
Increasing the threshold slightly modifies the shape of the
distribution: the height of the maximum increases and the curve is shifted to
lower $L$, i.e., the peaks fall more rapidly for higher $\nu$ .

	On the other hand, we can obtain the mean $L$ for the different models. 
For the case of FWHM=$10'$ and $A_{(N)}(10')=3\times 10^{-5}$, 
we find (in arcmin$^2$) $<L>=36.0$ and $38.7$ for $\Omega=0.1$ and $1$, 
respectively.
If we consider the same cases for $A_{(N)}(10')=0$ the corresponding mean $L$'s are
given by $106.9$ ($\Omega=0.1$) and $177.7$ ($\Omega=1$). 
Then, since the error in $<L>$ due to cosmic variance is expected to be very
small for the small angular scales considered, we can also use the mean values 
of $L$ to distinguish between the flat and open models.
In Table 3 the mean L's are given for the models considered.

	In order to measure the gaussian  curvature from a map obtained by an
experiment, the required pixel size would need to be approximately one fifth of the
typical curvature radius of the maxima. The two curvature radii for each peak
can be measured by a fit to a paraboloid centered on the maximum temperature.
The pixel size should be a compromise between having an appropriate number of 
pixels to perform the fit and remaining in the vicinity of the maximum.
In particular, if we want to test $\Omega$ values as low as $0.1$ for an 
angular resolution of $10'$ and $A_{(N)}(10')=0$ the required 
size should be $\approx 2'$.

\section {The distribution of eccentricities}

The distribution of peaks above the threshold $\nu $ with eccentricity
between $(\epsilon , \epsilon +d\epsilon )$, $p(\epsilon, >\nu)$, can be 
obtained from the following p.d.f.

\begin{eqnarray}
p(\epsilon, \nu ) & = &\frac{32(6)^{1/2}}{\pi }  e^{-\frac{1}{2}{\nu }^2}
{\epsilon }^3 \frac {
(1 - {\epsilon }^2)}{(2 - {\epsilon }^2)^{5}} \left< (H \pi)^{1/2}
e^{-G}\left(1 - \frac{1}{2}{\rm erfc}(H^{1/2}\gamma \nu s)\right)
\right.\nonumber \\
 & & [3H^2{(1 - {\gamma }^2)}^2 + 
6H^3{\gamma }^2(1 - {\gamma }^2){\nu }^2 + (H\gamma \nu )^4] +
 \nonumber \\
 & & \left. e^{-s^2{\gamma }^2{\nu }^2}s \left(5H^3\gamma {(1 - {\gamma 
}^2)}^2\nu + 
H^4(\gamma \nu )^3(1 - {\gamma }^2)\right)\right>,
\end{eqnarray}

\begin{equation}
H = \frac{(2 - {\epsilon }^2)^2}{(3 - 2{\gamma }^2){\epsilon }^4 + 
4(1 - {\epsilon }^2)},\ \ \ 
G = H\frac{(\gamma \nu {\epsilon }^2)^2}{(2 - {\epsilon }^2)^2}.
\end{equation}

We note that there is an error in the expression given by Bond and
Efstathiou(1987) for the conditional probability $P(e|\nu)$, where e is the
ellipticity related to $\epsilon$ by $\epsilon=2(e/(1+2e))^{1/2}$.
We have studied $p(\epsilon, >\nu)$ for different angular resolutions, 
noise amplitudes $A_{(N)}(10')$, thresholds and models. The main conclusion is that 
it would be difficult to
distinguish between the cosmological models based on the comparison of 
eccentricities. As a typical
example, in Figure 2, we represent the p.d.f 
$p(\epsilon, >\nu)$ 
for the angular resolution of $10'$, threshold $\nu = 3$ and no noise. The
introduction of some level of noise clearly makes things worse. Hence we can 
generically 
say that the eccentricity p.d.f. is not a good test for distinguishing
 between flat 
and
open models. 
On the other hand, Gurzadyan \& Kocharyan (1992, 1993) argue that mixing of 
photons in
a universe with negative curvature will produce elongated shapes as compared to 
the
flat case. All of our results lead to the opposite conclusion: the eccentricity 
p.d.f.
for flat and low-$\Omega$ universes show similar bell-shape (for thresholds 
$\nu =
3, 4$) with mean value $<\epsilon >\approx 0.7$ and almost the same dispersion.
Therefore, we conclude that the eccentricity is a bad discriminator of the
$\Omega$ parameter.

\section{Conclusions}

We have studied the distribution of peaks above a threshold $\nu $ using 
the mean
number and two local quantities: gaussian curvature and eccentricity. 
We have
considered a whole sky coverage, with angular resolutions 
of $5, 10, 20$ arcmin (antenna FWHM) and different levels of noise
$A_{(N)}(10')=(0,1,3)\times 10^{-5}$, and we have 
calculated the distribution of these quantities for
flat and open CDM models (with a
Harrison-Zel'dovich primordial spectrum). Our main conclusions are that the 
number of peaks
and the gaussian curvature are good discriminators of the geometry of
the universe,
whereas the eccentricity cannot be used to distinguish
between different $\Omega$ values. For thresholds $\nu = 3, 4$, these curves 
are indistinguishable
for flat and open models, and we disagree with Gurzadyan \& Kocharyan 
(1992, 1993) who
argue that mixing of photons in a space of negative curvature would  tend to elongate 
the spots in the CMB. On the other hand, an angular resolution of $10'$ is the
most appropiate to distinguish between low-$\Omega$ and flat models when 
noise is present.

\acknowledgements 
We would like to thank N. Sugiyama for providing us the radiation 
power spectrum and the referee E.L. Wright for his useful comments.
EMG and JLS acknowledge financial support from the Spanish DGICYT, 
project PB92-0434-C02-02. RBB acknowledges a Spanish M.E.C. Ph.D. 
scholarship. JS and LC have been supported in part by a grant from NASA. 
LC was also supported by a Fullbright Fellowship. We acknowledge 
financial support from the PECO contract of the EU ERBCIPDCT 940019.

\newpage

\figcaption{Logarithm of the number of peaks 
above a threshold $\nu$ versus the threshold, for 
different angular resolutions. The dashed and dotted lines correspond 
to open universes
($\Omega$ =0.1 and 0.3, respectively) and the solid one to a flat universe. 
Each set of three different lines corresponds to values of 
$A_{(N)}(10')=(3,1,0)\times 10^{-5}$
(from top to bottom).}

\figcaption{We plot in three of the figures the distribution of $L$ for
peaks above a threshold $\nu =3$ and several 
angular resolutions. The first, second and third set of 3 different lines
(from left to right)
corresponds to values of $A_{(N)}(10')=(3,1,0)\times 10^{-5}$, respectively. 
The p.d.f. p($\epsilon ,>\nu$ ) for a signal-dominated map ($A_{(N)}(10')=0$), 
$\nu$ =3 and FWHM=10$'$ is shown in the bottom right figure.
In the four plots the dashed, dotted and solid lines correspond to the cases
$\Omega$ =0.1,0.3 and 1, respectively.}

\end{document}